\documentclass[twocolumn,prc,amssymb, aps,superscriptaddress, 
amsmath]{revtex4}

\usepackage{color}
\usepackage{amsmath,mathtools,booktabs,
            microtype} 

\usepackage{graphicx}
\usepackage{dcolumn}
\usepackage{bm}
\usepackage{multirow}
\usepackage{times}

\usepackage{url}
\usepackage{tikz}
\usepackage[version=4]{mhchem}

\usetikzlibrary{arrows,decorations.pathmorphing}
\usetikzlibrary{arrows.meta}
\usetikzlibrary{calc}

\definecolor{violet}{HTML}{602969}
\definecolor{red}{HTML}{FC0009}
\definecolor{orange}{HTML}{FF6319}
\definecolor{green}{HTML}{00933C}
\definecolor{blue}{HTML}{0036A6}
\definecolor{yellow}{HTML}{FFBE00}
\definecolor{lightgrey}{HTML}{A7A9AC}

\newcommand{\be}{\begin{equation}}
\newcommand{\ee}{  \end{equation}}
\newcommand{\ba}{\begin{eqnarray}}
\newcommand{\ea}{  \end{eqnarray}}
\newcommand{\ve}{\varepsilon}

\begin{document}

\title{Random-Matrix Model for Thermalization}

\author{Hans A. \surname{Weidenm\"uller}}
\email{haw@mpi-hd.mpg.de}
\affiliation{Max-Planck-Institut f\"ur Kernphysik, Saupfercheckweg 1,
  D-69117 Heidelberg, Germany}

\date{\today}


\begin{abstract}
An isolated quantum system is said to thermalize if ${\rm Tr} (A
\rho(t)) \to {\rm Tr} (A \rho_{\rm eq})$ for time $t \to \infty$. Here
$\rho(t)$ is the time-dependent density matrix of the system,
$\rho_{\rm eq}$ is the time-independent density matrix that describes
statistical equilibrium, and $A$ is a Hermitean operator standing for
an observable. We show that for a system governed by a random-matrix
Hamiltonian (a member of the time-reversal invariant Gaussian
Orthogonal Ensemble (GOE) of random matrices of dimension $N$), all
functions ${\rm Tr} (A \rho(t))$ in the ensemble thermalize: For $N
\to \infty$ every such function tends to the value ${\rm Tr} (A
\rho_{\rm eq}(\infty)) + {\rm Tr} (A \rho(0)) g^2(t)$. Here $\rho_{\rm
  eq}(\infty)$ is the equilibrium density matrix at infinite
temperature. The oscillatory function $g(t)$ is the Fourier transform
of the average GOE level density and falls off as $1 / |t|$ for large
$t$. With $g(t) = g(-t)$, thermalization is symmetric in
time. Analogous results, including the symmetry in time of
thermalization, are derived for the time-reversal non-invariant
Gaussian Unitary Ensemble (GUE) of random matrices. Comparison with
the ``eigenstate thermalization hypothesis'' of Ref.~\cite{Sre99}
shows overall agreement but raises significant questions.

\end{abstract}

\maketitle

\section{Introduction}
\label{int}

Does an isolated quantum system thermalize? If so, under which
conditions does that happen? Is thermalization linked to time-reversal
invariance? These questions are important for the understanding of
fundamental aspects of quantum statistical mechanics. Thermalization
is seen as the quantum analogue of ergodicity in classical systems,
see the reviews in Refs.~\cite{Dal16, Deu18, Aba19}. Therefore,
thermalization continues to be discussed intensely until
today~\cite{Ued20, Shi21}. Thermalization cannot occur in quantum
systems that are integrable or show many-body localization, and the
discussion is focused on chaotic quantum systems. In an isolated
system, thermalization cannot be investigated experimentally by
definition. In ultracold atoms~\cite{Ued20}, for instance, the
difficulty is circumvented by opening the system after some time and
taking data.

In Refs.~\cite{Dal16, Aba19}, thermalization is defined in terms of
the time-dependent function ${\rm Tr} (A \rho(t))$. Here $\rho(t)$ is
the time-dependent density matrix of an isolated quantum system, and
$A$ is a Hermitean operator representing an observable used to test
thermalization. The system is said to thermalize if for $t \to
\infty$, ${\rm Tr} (A \rho(t))$ tends asymptotically to ${\rm Tr} (A
\rho_{\rm eq})$ where $\rho_{\rm eq}$ is the time-independent density
matrix of statistical equilibrium. Thus, thermalization provides a
test of the general hypothesis that in the long-time limit,
expectation values of operators tend to their equilibrium values.

A central role in the discussion plays the ``eigenstate thermalization
hypothesis'' introduced by Srednicki~\cite{Sre99}. In the basis of
eigenstates $\alpha, \beta, \ldots$ of the chaotic Hamiltonian with
energy eigenvalues $E_\alpha, E_\beta, \ldots$, the operator $A$ is
represented by the matrix $A_{\alpha \beta}$. The hypothesis
postulates that $A_{\alpha \beta}$ has the form
\ba
\label{i1}
A_{\alpha \beta} = {\cal A}(E) \delta_{\alpha \beta} + \exp \{ - S(E)/2 \}
F(E, \omega) R_{\alpha \beta} \ .
\ea
Here ${\cal A}(E)$ and $F(E, \omega)$ are smooth functions of $E =
(1/2)(E_\alpha + E_\beta)$ and $\omega = E_\alpha - E_\beta$, $S(E)$ is
the thermodynamic entropy at energy $E$, and $R_{\alpha \beta}$ is a
zero-centered Gaussian-distributed random variable with unit variance.
On the basis of Eq.~(\ref{i1}) it is argued that in the semiclassical
regime of high excitation energy, thermalization is generic for
chaotic quantum systems~\cite{Sre99}.

That is an amazing assertion, for the following reason. The state of a
quantum system is characterized by the Hermitean time-independent
statistical operator $\Pi$ that describes the distribution of the
system over the states in Hilbert space. With $H$ the Hamiltonian
governing the system, the time-dependent density matrix $\rho(t)$ is
given by
\ba
\label{i2}
\rho(t) = \exp \{ - i H t / \hbar \} \Pi \exp \{ i H t / \hbar \} \ .  
\ea
In the eigenstate representation of $H$, the diagonal matrix elements
$\rho_{\alpha \alpha}$ of $\rho(t)$ are independent of time, and the
non-diagonal elements $\rho_{\alpha \beta}$ with $\alpha \neq \beta$
carry the time-dependent phase factors $\exp \{ - i (E_\alpha -
E_\beta) t / \hbar \} $. These oscillate periodically in time. The
operator $A$ is independent of time. Therefore, the time-dependent
function ${\rm Tr} (A \rho(t))$ has a very simple structure,
irrespective of whether the system is chaotic or not. How can that
simplicity be reconciled with the statement that ${\rm Tr} (A
\rho(t))$ thermalizes, i.e., tends for time $t \to \infty$ to the
limit ${\rm Tr} (A \rho_{\rm eq})$? Why do chaotic systems thermalize
while others do not? Eq.~(\ref{i1}) indicates that the answer given in
Ref.~\cite{Sre99} is a probabilistic one. The random fluctuations of a
chaotic quantum system encapsulated in the last term of Eq.~(\ref{i1})
are used in Ref.~\cite{Sre99} to show that with overwhelming
probability, ${\rm Tr} (A \rho(t))$ tends to ${\rm Tr} (A \rho_{\rm
  eq})$ in the long-time limit.

In the present paper we investigate thermalization from a different
point of view. The random fluctuations of eigenvalues and
eigenfunctions of chaotic quantum systems agree locally in energy with
random-matrix predictions~\cite{Boh84}.  That fact has motivated us to
investigate the thermalization of a quantum system for which the
Hamiltonian $H$ is a random matrix. Throughout most of the paper, $H$
is taken to be a member of the time-reversal invariant Gaussian
Orthogonal Ensemble (GOE) of random matrices of dimension $N \gg 1$
while Section~\ref{uni} is devoted to the Gaussian Unitary Ensemble
(GUE) that breaks time-reversal invariance. By definition, $H$ is
paradigmatic for the fluctuation properties that characterize chaotic
quantum systems. Clearly, however, the structureless random-matrix
Hamiltonian $H$ which treats all states of the system on an equal
footing, cannot represent a real chaotic quantum system with all its
information content. We are going to show that, nevertheless, our
study offers useful insights into the conditions for and
characteristic properties of thermalization. We investigate the
conditions under which ${\rm Tr} (A \rho(t))$ thermalizes for a GOE
Hamiltonian. We show that under the same conditions, ${\rm Tr} (A
\rho(t))$ thermalizes in very similar form also for a GUE
Hamiltonian. In comparing our results with those of Ref.~\cite{Sre99}
we illuminate the difference between the two approaches.

Each realization of the random-matrix Hamiltonian $H$ defined in
Section~\ref{def} produces another time-dependent test function ${\rm
  Tr} (A \rho(t))$. Jointly, these test functions form a ``stochastic
process'', i.e., a random ensemble of time-dependent functions. We
determine analytically mean value (Section~\ref{ens}) and correlation
function (Section~\ref{cor}) of that stochastic process in the limit
$N \gg 1$ of large matrix dimension. We calculate terms of order zero
in $N$ as well as corrections of order $N^{- 1}$. We thereby explore
the conditions under which every member ${\rm Tr} (A \rho(t))$ of the
stochastic process thermalizes. In Section~\ref{uni} we investigate
thermaliztion for the GUE in the same manner. In Section~\ref{eig} we
compare our approach and our results with those obtained in the
framwork of the eigenstate thermalization
hypothesis. Section~\ref{dis} contains a discussion and the
conclusions.


\section{Definitions}
\label{def}

All operators act in a Hilbert space of dimension $N \gg 1$, spanned
by orthonormal basis states labeled $(\mu, \nu, \ldots)$. The function
${\rm Tr} (A \rho(t))$ depends on time $t$ via the density matrix
$\rho(t)$ defined in Eq.~(\ref{i2}). The Hermitean time-independent
statistical operator $\Pi$ describes the distribution of the system
over the states in Hilbert space, has eigenvalues $\pi_\kappa$,
orthonormal eigenfunctions $| \kappa \rangle$ with $\kappa = 1,
\ldots, N$, and is given by
\ba
\label{d2}
\Pi_{\mu \nu} = \sum_\kappa \langle \mu | \kappa \rangle \pi_\kappa
\langle \kappa | \nu \rangle \ .
\ea
The real eigenvalues $\pi_\kappa$ obey $0 \leq \pi_\kappa \leq 1$ for
all $\kappa$. The matrix $\Pi$ has unit trace, $\sum_\kappa \pi_\kappa
= 1$. It is capable of accounting for very different physical
situations. If, for instance, $\pi_\kappa = 1 / N$ for all $\kappa$,
we have $\Pi_{\mu \nu} = (1 / N) \delta_{\mu \nu} = (\Pi_{\rm
  eq})_{\mu \nu}$. That operator describes statistical equilibrium at
infinite temperature where the Boltzmann factor is equal to unity and
all states have equal occupation probability. If $\pi_\kappa = 2 / N$
for $\kappa \leq N / 2$ and $\pi_\kappa = 0$ for $\kappa > N / 2$, the
system is far from equilibrium. In both cases, however, ${\rm Tr}
(\Pi)^2$ is of order $1 / N$ and, thus, small compared to $[{\rm Tr}
  \Pi]^2 = 1$. A very different case is the one where $\pi_\kappa = 1$
for $\kappa = 1$ and $\pi_\kappa = 0$ otherwise. Then ${\rm Tr}
(\Pi)^2 = 1$ and equal to $[{\rm Tr} \Pi]^2 = 1$. Cases intermediate
between these extremes also exist. If, for instance, $\pi_\kappa =
N^{- \alpha}$ for $0 < \alpha < 1$ and for all $\kappa \leq
N^{\alpha}$ and zero otherwise, we have ${\rm Tr} \Pi = 1$ and ${\rm
  Tr} (\Pi)^2 = N^{- \alpha}$. These cases are distinguished by the
rank $k$ of $\Pi$. In the first, second, third, and fourth example we
have, respectively, $k = N$, $k = N / 2$, $k = 1$, and $k =
N^{\alpha}$, and ${\rm Tr} (\Pi)^2$ is of order $1 / k$. In discussing
thermalization we keep in mind these different possibilities.

The operator $\Pi$ is independent of time. The time evolution of the
density matrix in Eq.~(\ref{i2}) is entirely determined by the unitary
operators $\exp \{ \pm i H t / \hbar \}$. (We assume that $\Pi \neq
\Pi_{\rm eq}$ as otherwise $\rho(t) = \Pi_{\rm eq}$ is independent of
time, and the system remains in statistical equilibrium forever.)
Except for Section~\ref{uni}, the Hamiltonian $H$ is a member of the
GOE. The matrix elements $H_{\mu \nu}$ are real zero-centered Gaussian
random variables with second moments
\ba
\big\langle H_{\mu \nu} H_{\mu' \nu'} \big\rangle = \frac{\lambda^2}{N}
\big( \delta_{\mu \mu'} \delta_{\nu \nu'} + \delta_{\mu \nu'}\delta_{\nu \mu'}
\big) \ .
\label{d3}
\ea
Big angular brackets indicate the ensemble average, and $\lambda$
defines the width of the spectrum. In terms of its eigenvalues
$E_\alpha$ and eigenfunctions $O_{\mu \alpha}$ with $\alpha = 1,
\ldots, N$, the Hamiltonian is given by
\ba
\label{d4}
H_{\mu \nu} = \sum_\alpha O_{\mu \alpha} E_\alpha O_{\nu \alpha} \ .
\ea
For $N~\gg~1$ the matrix elements $O_{\mu \alpha}$ are real
zero-centered Gaussian random variables~\cite{Por56} with second
moments
\ba
\label{d5}
\big\langle O_{\mu \alpha} O_{\mu' \alpha'} \big\rangle = \frac{1}{N}
\delta_{\mu \mu'} \delta_{\alpha \alpha'} \ .
\ea
The factor $(1 / N)$ accounts for normalization. The eigenvalues
$E_\alpha$ obey Wigner-Dyson statistics. Eigenvectors and eigenvalues
are statistically uncorrelated~\cite{Meh04}. The unitary
time-evolution operator $U = \exp \{ - i H t / \hbar \}$ is given by
\ba
\label{d6}
U_{\mu \nu}(t) = \sum_\alpha O_{\mu \alpha} \exp \{ - i E_\alpha t / \hbar \}
O_{\nu \alpha} \ .
\ea
The operator $U$ carries the random variables $O_{\mu \alpha}$ and
$E_\alpha$ and is, therefore, a matrix-valued stochastic process.

In analogy to the operator $\Pi$ in Eq.~(\ref{d2}), the operator $A$
with eigenvalues $A_j$ and orthonormal eigenfunctions $| j \rangle$ is
written as
\ba
\label{d7}
A_{\mu \nu} &=& \sum_{j} \langle \mu | j \rangle A_j \langle j | \nu
\rangle \ ,
\ea
and we have
\ba
\label{d8}
{\rm Tr} (A \rho(t)) = {\rm Tr} [A U(t) \Pi U^\dag(t)] \ . 
\ea
The operators $A$ and $\Pi$ are independent of time and not
statistical. Both are Hermitean. Therefore, ${\rm Tr} (A \rho(t))^* =
{\rm Tr} (A \rho(t))$ is real. The operator $U(t)$ defined in
Eq.~(\ref{d6}) is both time-dependent and stochastic and so is,
therefore, also ${\rm Tr} (A \rho(t))$. We investigate thermaliztion
by calculating the ensemble average and the correlation function of
${\rm Tr} (A \rho(t))$. To that end we calculate in Appendix~1 the low
moments of $U(t)$ and in Appendix~2 the time-dependent functions that
control thermalization.

\section{Ensemble Average}
\label{ens}

We average separately over matrix elements $O_{\mu \alpha}$ and
eigenvalues $E_\alpha$. We begin with the matrix elements. The average
is indicated by an index $O$ on the big angular brackets. We use
Eq.~(\ref{d8}) and separate uncorrelated and correlated factors $U(t)$
to write
\ba
\label{e1}
&& \big\langle {\rm Tr} (A \rho(t)) \big\rangle_O = {\rm Tr} \big[ A
  \big\langle U(t) \big\rangle_O \Pi \big\langle
  U^\dag(t) \big\rangle_O \big] \nonumber \\
&& \qquad + {\rm Tr} \big[ A \big\langle U(t) \Pi 
  U^\dag(t) \big\rangle_{O, {\rm corr}} \big] \ .
\ea
We use the averages given in Eqs.~(\ref{a3}) and (\ref{a5}) of
Appendix~1 and ${\rm Tr} (\Pi) = 1$ to obtain
\ba
\label{e2}
&& \big\langle {\rm Tr} (A \rho(t)) \big\rangle = \big\langle
|f(t)|^2 \big\rangle {\rm Tr} (A \Pi) + \frac{1}{N} {\rm Tr} (A)
\nonumber \\
&& \qquad + \frac{1}{N} {\rm Tr} (A^T \Pi) \ .
\ea
Here
\ba
\label{e3}
f(t) = \frac{1}{N} \sum_\alpha \exp \{ - i E_\alpha t / \hbar \} \ . 
\ea
The upper index $T$ in expression~(\ref{e2}) denotes the transpose,
the angular brackets denote the remaining average over the
eigenvalues. The function $f(t)$ is obviously of order zero in $N$ and
so are the first and the second term on the right-hand side of
Eq.~(\ref{e2}) whereas the third term is of order $N^{- 1}$. That can
be verified using the four forms of $\Pi$ listed below
Eq.~(\ref{d2}). The second term on the right-hand side of
Eq.~(\ref{e2}) can be written as ${\rm Tr} (A \rho_{\rm
  eq}(\infty))$. Here $(\rho_{\rm eq}(\infty))_{\alpha \beta} = (1 /
N) \delta_{\alpha \beta} = (\Pi_{\rm eq})_{\alpha \beta}$ is the
time-independent density matrix that describes statistical equilibrium
at infinite temperature. If $\big\langle | f(t) |^2 \big\rangle$ tends
to zero for large time, Eq.~(\ref{e2}) implies thermalization to
leading order in $N$ for the ensemble average of ${\rm Tr} (A
\rho(t))$. The rate of thermalization depends only on $\big\langle |
f(t) |^2 \big\rangle$, i.e., on the eigenvalues $E_\alpha$ of $H$, and
is independent of the test operator $A$ and of the statistical
operator $\Pi$.

We note that at time $t = 0$, we have $U(0) = 1$ and, from
Eq.~(\ref{d8}), ${\rm Tr} (A \rho(0)) = {\rm Tr} (A \Pi)$.  That
result differs from the limit $t \to 0^\pm$ of
expression~(\ref{e2}). With $f(0) = 1$ the limit is given by $(1 / N)
{\rm Tr} (A) + {\rm Tr} (A \Pi) + (1 / N) {\rm Tr} (A^T \Pi)$. Thus,
$t = 0$ is a singular point in the time evolution of ${\rm Tr} (A
\rho(t))$.

We express $\big\langle |f(t)|^2 \big \rangle$ as the sum of the
uncorrelated and the correlated parts,
\ba
\label{e4}
\big\langle |f(t)|^2 \big \rangle = |\big\langle f(t) \big\rangle|^2
+ \big\langle |f(t)|^2 \big\rangle_{\rm corr} \ .
\ea
Both parts are calculated in Appendix~2. The resulting expressions
involve the time scales
\ba
\label{e9a}
\tau_\lambda = \hbar / (2 \lambda) \ , \ \tau_d = \hbar / d \ .
\ea
Here $2 \lambda$ defines the width of the average GOE spectrum, and
$d$ is the average GOE level spacing. At the center $E = 0$ of the
spectrum, $d$ has the value $d = \pi \lambda / N$. Since $2 \lambda$
is the largest and $d$ is the smallest natural GOE energy scale,
$\tau_\lambda$ is the shortest and $\tau_d$ is the largest natural GOE
time scale. The ratio is $\tau_\lambda / \tau_d = \pi / (2 N) \ll 1$.

The uncorrelated part
\ba
\label{e10}
\langle f(t) \rangle = g(t / \tau_\lambda) = \frac{4}{\pi} \int_0^1
{\rm d} x \ \sqrt{1 - x^2} \cos( x t / \tau_\lambda)
\ea
is proportional to the Fourier transform of the average GOE level
density in Eq.~(\ref{e7}). The oscillatory function $g(t /
\tau_\lambda)$, given in detail in Appendix~2, obeys $g(0) = 1$, is
symmetric in time so that $g(t / \tau_\lambda) = g(- t /
\tau_\lambda)$, and for large time $|t|$ is bounded by $|g(t /
\tau_\lambda)| \leq 2 \lambda / |t|$.

The correlated part
\ba
\label{e16}
\langle |f(t)|^2 \rangle_{\rm corr} = \frac{4}{3 \pi N}
\int_{- \infty}^{+ \infty} {\rm d} y \ Y_2(y) \cos (y t / \tau_d) \} \ .
\ea
is proportional to the Fourier transform of the GOE two-level
correlation function $Y_2(y)$ defined in Eq.~(\ref{e12a}). The factor
$1 / N$ is due to the fact that $Y_2(y)$ is essentially confined to
values $y \ll N$ of the argument.

In summary we have shown that
\ba
\label{e17}
&& \big\langle {\rm Tr} (A \rho(t)) \big\rangle = \frac{1}{N} {\rm Tr}
(A) \\
&& + {\rm Tr} (A \Pi) \big( g^2(t / \tau_\lambda) + \big\langle |f(t)|^2
\big\rangle_{\rm corr}\big) + \frac{1}{N} {\rm Tr} (A^T \Pi) \ . \nonumber
\ea
The first term on the right-hand side represents the statistical
equilibrium value ${\rm Tr} (A \rho_{\rm eq}(\infty))$ of ${\rm Tr} (A
\rho(t))$. The second term describes thermalization. Since both $g(t /
\tau_\lambda)$ and $\langle |f(t)|^2 \rangle_{\rm corr}$ are symmetric
under the operation $t \to - t$, thermalization is symmetric in
time. The function $g^2$ falls off with time as $4 \tau^2_\lambda /
t^2$ until it reaches the level of the statistical fluctuations given
by $\big\langle |f(t)|^2 \big\rangle_{\rm corr}$ in
Eq.~(\ref{e16}). Since the fluctuations are of order $1 / N \approx
\tau_\lambda / \tau_d$, that happens when $g^2 \approx 1 / N$ or, with
$|g| \approx \tau_\lambda / t$, at time $t \approx \sqrt{\tau_\lambda
  \tau_d}$ intermediate between the shortest time $\tau_\lambda$ and
the longest time $\tau_d$. Beyond that time, the fluctuations take
over. These are governed by $\tau_d$ and, thus, are extremely
slow. The last term in Eq.~(\ref{e17}) is of order $(1 / N)$. It
provides an offset for the fluctuations which prevents for finite $N$
the test function ${\rm Tr} (A \rho(t))$ from ever thermalizing
completely. The dependence on time both of $g$ and of the correlated
part in Eq.~(\ref{e16}) is determined entirely by the statistical
properties of the GOE and is independent of the test operator $A$ and
of the statistical operator $\Pi$. These two operators determine the
values of the last two traces in Eq.~(\ref{e17}) and, thus, initial
and final values of thermalization.

The phase factors $\exp \{ \pm i E_\alpha t / \hbar \}$ in
expression~(\ref{e1}) represent solutions of the Schr{\"o}dinger
equation. That equation is invariant under a time shift $t \to t +
t_0$. We are free to use that shift in expression~(\ref{e1}) and, in
consequence, in $g(t / \tau_\lambda)$. The resulting function $g((t +
t_0) / \tau_\lambda)$ is a valid alternative to $g(t / \tau_\lambda)$.
The maximum is at $t = - t_0$. Depending on the sign of $t_0$, the
maximum occurs at negative or positive values of $t$. In the latter
case thermalization has the unexpected feature that, starting at $t =
0$, $g^2$ first grows (with oscillations) with increasing $t$, reaches
the absolute maximum $g = 1$ at $t = t_0$, and decreases thereafter.

As mentioned in the Introduction, thermalization is
understood~\cite{Dal16} as a test of the hypothesis that in the
long-time limit, expectation values of operators tend to their
equilibrium values. Thermalization does not imply that the density
matrix $\rho(t)$ itself thermalizes, i.e., tends to the
time-independent statistical equilibrium distribution. That is
demonstrated explicitly by the freedom to shift time $t \to t + t_0$
which shows that the onset in time of thermalization can be chosen
arbitrarily. Choosing $t_0 > 0$ shows that $\rho(t)$ does not
thermalize in the time interval $0 \leq t \leq t_0$.

\section{Correlation Function}
\label{cor}

As a measure of the size of the fluctuations of individual members
${\rm Tr} (A \rho(t))$ around the ensemble average in Eq.~(\ref{e17})
we calculate the correlation function
\ba
\label{c1}
\big\langle {\rm Tr} (A \rho(t)) {\rm Tr} (A \rho(t)) \big\rangle
- \big\langle {\rm Tr} (A \rho(t)) \big\rangle \big\langle
{\rm Tr} (A \rho(t)) \big\rangle \ .
\ea
We first perform the average over the matrix elements $O_{\mu
  \alpha}$. Following Appendix~1 we distinguish contributions due to
the first moment of $U$ and to the totally correlated parts of the
second, third, and fourth moment of $U(t)$. Nonvanishing contributions
to expression~(\ref{c1}) are due to the following possibilities:
(i)~In each trace in the first term of expression~(\ref{c1}), each of
the two factors $U$ is averaged individually. Then the only
non-vanishing contribution to the correlation function~(\ref{c1}) is
due to the correlations between eigenvalues. (ii) In each trace one
factor $U$ is part of the correlated part of the second moment, the
other factor is replaced by its average (four cases); (iii) in each
trace each factor $U$ is paired with a factor $U$ in the other trace
to form the correlated part of the second moment (two cases); (iv) two
factors $U$ in one trace combine with one factor $U$ in the second
trace to form the correlated part of the third moment, the remaining
factor $U$ is averaged separately (four cases); (v) the four factors
$U$ form the correlated part of the fourth moment (one case).

For case~(i) Eq.~(\ref{a3}) shows that the contribution to the
correlation function is given by
\ba
\label{c2}
\big( \big\langle |f(t)|^2 |f(t)|^2 \big\rangle - \big\langle
|f(t)|^2 \big\rangle \big\langle |f(t)|^2 \big\rangle \big) \big(
{\rm Tr} (A \Pi) \big)^2 \ .
\ea
Here ${\rm Tr} (A \Pi)$ is of order zero in $N$. The correlation
function of $|f(t)|^2$ involves the GOE two-point correlation function
in Eq.~(\ref{e16}) as well as the GOE three-point and four-point
correlation functions. As pointed out below Eq.~(\ref{e16}) and in
Appendix~2, the factor $1 / N$ in Eq.~(\ref{e16}) is due to the fact
that the integration variable $y$ is essentially confined to values of
order unity. In the GOE three-point (four-point) functions, two
(three) integration variables are similarly confined. That gives rise,
respectively, to factors $1 / N^2$ and $1 / N^3$ multiplying these
functions. In evaluating the correlation function in
expression~(\ref{c2}) we confine ourselves to contributions of order
$1 / N$. These are due to the two-point function and given by
\ba
\label{c3}
&& \big( \big\langle f(t) \big\rangle \big\langle f(t) \big\rangle
\big\langle f^*(t) f^*(t) \big\rangle_{\rm corr} + c.c. \big)
\nonumber \\
&& \ \ + \big( \big\langle f(t) \big\rangle \big\langle f^*(t)
\big\rangle \big\langle f^*(t) f(t) \big\rangle_{\rm corr} + c.c.
\big) \ .
\ea
Here $\big\langle f(t) \big\rangle$ is given in Eq.~(\ref{e10}) and
the two-point correlation function is given in Eq.~(\ref{e16}). All
terms are of order $1 / N$ and are of second order in $g$ and, thus,
disappear with $t^{- 2}$ for large $|t|$.

For case~(ii) the results carry the factor $1 / N$, factors $f(2 t)
(f^*(t))^2$ or $|f(t)|^2$, and traces of the form ${\rm Tr} (A^2
\Pi^2)$, ${\rm Tr} (\Pi A A^T \Pi^T)$, or ${\rm Tr} (\Pi A)^2$. To
leading order in $N$, all time-dependent factors are proportional to
$g^k$ with $k \geq 2$ and tend to zero for large $|t|$. The factor
${\rm Tr} (A^2 \Pi^2)$, representative for all terms, is of order $1 /
N$ if the rank of $\Pi$ is of order $N$ and is of order unity if the
rank of $\Pi$ is of order unity. That changes if we request that $A$
be almost diagonal, with diagonal elements $A_j$ differing by terms of
order $1 / N$, or that
\ba
\label{c5}
\big( {\rm Tr} (A) \big)^2 \gg {\rm Tr} (A^2) \ ,
\ea
the inequality indicating a ratio of order $1 / N$. The
postulate~(\ref{c5}) resembles the eigenstate thermalization
hypothesis of Eq.~(\ref{i1}). It does not, however, invoke any
statistical assumption on $A$. If condition~(\ref{c5}) is met, the
term $(1 / N) {\rm Tr} (A^2 \Pi^2)$ is of order $1 / N$ if the rank of
$\Pi$ is of order unity and is considerably smaller than that if the
rank of $\Pi$ is of order $N$. That same statement holds for the other
contributions to case~(ii).

For case~(iii) the results carry a factor $1 / N^2$, a factor unity or
$|f(2t)|^2$, and factors $\sum_{\mu \nu} A_{\mu \nu}^2 \Pi_{\mu
  \nu}^2$, ${\rm Tr} (A^2) {\rm Tr} (\Pi^2)$, or $[{\rm Tr} (A
  \Pi)]^2$. All these terms are small of order $1 / N$ or smaller,
without any constraints on $A$. For case (iv) the results carry a
factor $(1 / N)^2 |f(t)|^2$ and factors ${\rm Tr} (\Pi A A^T)$, ${\rm
  Tr} (A) {\rm Tr} (A \Pi \Pi^T)$. These are small of order $1 / N$.
For case (v) the results carry a factor $1 / N^3$ and factors ${\rm
  Tr} (A^2)$, ${\rm Tr} (A \Pi)$ ${\rm Tr} (A)$, or ${\rm Tr} (\Pi^2)$
$[{\rm Tr} (A)]^2$. These are small of order $1 / N$.

We have, thus, shown that all terms in the correlation
function~(\ref{c1}) are of order $1 / N$ or smaller. Without any
constraints imposed on $A$ that statement holds if the rank of $\Pi$
is of order $N$. If, on the other hand, the rank of $\Pi$ is of order
unity, the statement holds provided that $A$ obeys the
constraint~(\ref{c5}).

Summarizing the results of the previous and of the present Section, we
have shown that for matrix dimension $N \to \infty$, all members ${\rm
  Tr} (A \rho(t))$ of the stochastic process defined by the GOE
Hamiltonian $H$ tend to a common limit,
\ba
\label{s1}
{\rm Tr} (A \rho(t)) \to \frac{1}{N} {\rm Tr} (A) + {\rm Tr} (A \Pi)
\ g^2(t / \tau_\lambda) \ .
\ea
The first term on the right-hand side represents the statistical
equilibrium value ${\rm Tr} (A \rho_{\rm eq}(\infty))$ of ${\rm Tr} (A
\rho(t))$ at infinite temperature. The oscillatory function $g^2$ has
its maximum $g^2 = 1$ at $t = 0$, is bounded for large $|t|$ from
above by $(2 \tau_\lambda / t)^2$ and, thus, tends to zero for large
time $|t|$. Thus, each of the functions ${\rm Tr} (A \rho(t))$
thermalizes. Since $g(t / \tau_\lambda) = g(- t / \tau_\lambda)$,
thermalization is symmetric in time. The result~(\ref{s1}) holds in
full generality if the rank of $\Pi$ is of order $N$ and with the
constraint~(\ref{c5}) on $A$ if the rank of $\Pi$ is of order unity.

\section{Gaussian Unitary Ensemble}
\label{uni}

The symmetry in time of the thermaliztion process in Eq.~(\ref{s1}) is
evident already in Eq.~(\ref{e2}). Since $f(t)^* = f(- t)$, the
function $\big\langle |f(t)|^2 \big\rangle$ obeys $\big\langle
|f(t)|^2 \big\rangle = \big\langle |f(- t)|^2 \big\rangle$. To test
whether that symmetry is caused by the time-reversal invariance of the
GOE, we study thermalization for the Gaussian Unitary Ensemble (GUE)
of random matrices. That ensemble is not invariant under time
reversal. The GUE consists of Hermitean matrices $H_U$ of dimension $N
\gg 1$. The elements $H_{U; \mu \nu}$ are complex zero-centered
Gaussian random variables with second moments
\ba
\label{u1}
\big\langle H_{U; \mu \nu} H_{U; \mu' \nu'} \big\rangle =
\frac{\lambda^2_U}{N} \delta_{\mu \nu'} \delta_{\nu \mu'} \ .
\ea
Here $\lambda_U$ defines the width of the GUE spectrum. In terms of
its eigenvalues $E_{U; \alpha}$ and eigenfunctions ${\cal U}_{\mu
  \alpha}$ with $\alpha = 1, \ldots, N$, the Hamiltonian matrix $H_U$
is given by
\ba
\label{u2}
H_{U; \mu \nu} = \sum_\alpha {\cal U}_{\mu \alpha} E_{U; \alpha} \
{\cal U}^*_{\nu \alpha} \ .
\ea
The elements ${\cal U}_{\mu \alpha}$ of the unitary matrix ${\cal U}$
are complex zero-centered random variables that carry random
phases. Therefore, $\big\langle {\cal U}_{\mu \alpha} {\cal U}_{\mu'
  \alpha'} \big\rangle = 0$ while
\ba
\label{u3}
\big\langle {\cal U}_{\mu \alpha} {\cal U}^*_{\mu' \alpha'} \big\rangle
= \frac{1}{N} \delta_{\alpha \alpha'} \delta_{\mu \mu'} \ .
\ea
The eigenvalues $E_{U; \alpha}$ obey GUE Wigner-Dyson
statistics. Eigenvectors and eigenvalues are statistically
uncorrelated~\cite{Meh04}. The unitary time-evolution operator $U_U(t)
= \exp \{ - i H_U t / \hbar \}$ is given by
\ba
\label{u4}
U_{U; \mu \nu}(t) = \sum_\alpha {\cal U}_{\mu \alpha} \exp \{ - i
E_{U; \alpha} t / \hbar \} \ {\cal U}^*_{\nu \alpha} \ .
\ea
With the replacement $U(t) \to U_U(t)$, the stochastic process ${\rm
  Tr} (A \rho(t))$ is given by Eq.~(\ref{d8}). We average ${\rm Tr} (A
\rho(t))$ first over the eigenvectors ${\cal U}_{\mu \alpha}$. That is
done by contracting each of the two factors ${\cal U}_{\mu \alpha}$
with one of the two factors ${\cal U}^*_{\mu \alpha}$ and use of
Eq.~(\ref{u3}). There are two contraction patterns. The result is
\ba
\label{u5}
\big\langle {\rm Tr} (A \rho(t)) \big\rangle = \frac{1}{N} {\rm Tr}
(A) + \big\langle |f_U(t)|^2 \big\rangle {\rm Tr} (A \Pi) \ .
\ea
Here $f_U(t)$ is defined as in Eq.~(\ref{e3}) except for the
replacement $E_\alpha \to E_{U; \alpha}$. Proceeding as in Appendix~2
we see that $\big\langle f_U(t) \big\rangle$ is proportional to the
Fourier transform of the average GUE level density, and that the
correlated part $\big\langle |f_U(t)|^2 \big\rangle_{\rm corr}$ is
proportional to the Fourier transform of the GUE level correlation
function $Y_2(r) = \sin^2(\pi r) / (\pi r)^2$~\cite{Meh04}. The
avergage level densities of the GOE and of the GUE
coincide~\cite{Meh04}. Thus, $\big\langle f_U(t) \big\rangle$ equals
$g(t / \tau_\lambda)$ defined in Eq.~(\ref{e10}), whereas $\big\langle
|f_U(t)|^2 \big\rangle_{\rm corr}$ is obtained from $\big\langle
|f(t)|^2 \big\rangle_{\rm corr}$ in Eq.~(\ref{e16}) by replacing the
GOE two-level correlation function by its GUE counterpart. The steps
in Section~\ref{cor} can be followed to show that the GUE analogue to
the correlation function in Eq.~(\ref{c1}) vanishes for $N \to
\infty$. Thus, thermalization is very similar for the GUE and for the
GOE. Since $f^*_U(t) = f_U(- t)$ we have $\big\langle |f_U(t)|^2
\big\rangle = \big\langle |f_U(- t)|^2 \big\rangle$. The time
dependence of thermalization for the GUE is symmetric about $t =
0$. We conclude that the symmetry of thermalization with respect to
the replacement $t \to - t$ is not a consequence of time-reversal
invariance. It follows from the fact that $\rho(t)$ is Hermitean. That
causes $\rho(t)$ to have the forms $U(t) \Pi U^\dag(t)$ for the GOE
and $U_U(t) \Pi U^\dag_U(t)$ for the GUE. These forms, in turn, cause
the appearance of the time-symmetric factors $\big\langle |f(t)|^2
\big\rangle$ in Eq.~(\ref{s1}) and $\big\langle |f_U(t)|^2
\big\rangle$ in Eq.~(\ref{u5}).

\section{The eigenstate thermalization hypothesis}
\label{eig}

We compare our procedure and results with those of Ref.~\cite{Sre99}.
In the eigenbasis of the Hamiltonian $H$ with eigenvectors $| \alpha
\rangle$ and eigenvalues $E_\alpha$ we write Eq.~(\ref{d7}) as
\ba
\label{ei0}
{\rm Tr} (A \rho(t)) = \sum_{\alpha \beta} A_{\alpha \beta} \Pi_{\beta \alpha}
\exp \{ i (E_\alpha - E_\beta) t / \hbar \} \ .      
\ea
Here $A_{\alpha \beta} = \langle \alpha | A | \beta \rangle$ and
$\Pi_{\beta \alpha} = \langle \beta | \Pi | \alpha
\rangle$. Eq.~(\ref{ei0}) is completely general, i.e., holds for every
Hamiltonian, and is not restricted to the model defined in
Section~\ref{def}. In Ref.~\cite{Sre99} it is assumed that the system
is in a single state $| \psi(t) \rangle = \sum_\alpha c_\alpha \exp \{
- i E_\alpha t / \hbar \} | \alpha \rangle$. That is equivalent to
assuming that $\Pi$ has the special form
\ba
\label{ei01}
\Pi_{\beta \alpha} = c_\beta c^*_\alpha \ ,
\ea
i.e., that $\Pi$ has rank one. Moreover, it is assumed that $A_{\alpha
  \beta}$ has the form of Eq.~(\ref{i1}), with $R_{\alpha \beta}$ a
zero-centered Gaussian random variable.

We interpret Eq.~(\ref{i1}) and especially the occurrence of the
random variable $R_{\alpha \beta}$ as due to the fact that in a
chaotic quantum system, eigenfunctions and eigenvalues fluctuate
randomly~\cite{Boh84}. In the eigenstate representation of the
Hamiltonian, the matrix elements $A_{\alpha \beta}$ of the operator
$A$ are then random variables. These are parametrized in the form of
Eq.~(\ref{i1}).  Consistency would seem to require that the elements
$\Pi_{\beta \alpha}$ of the operator $\Pi$ in Eq.~(\ref{ei01}) and the
eigenvalues $E_\alpha$ in Eq.~(\ref{ei0}) are treated as random
variables, too. That is not done in Ref.~\cite{Sre99}. The only random
element in the approach of Ref.~\cite{Sre99} is the variable
$R_{\alpha \beta}$. Changing that and taking, in addition to
Eq,~(\ref{i1}), the energies $E_\alpha$ and the coefficients
$c_\alpha$ as random variables, would alter the results of
Ref.~\cite{Sre99}. We show that explicitly below Eq.~(\ref{ei5}) for
the function $C_{\rm eth}(t)$.

Aside from using the eigenstate thermalization hypothesis
Eq.~(\ref{i1}), the approach of Ref.~\cite{Sre99} investigates
thermalization of generic chaotic quantum systems with the help of
long-time averages of powers of ${\rm Tr} (A \rho(t))$ defined for $m
= 1, 2, \ldots$ as
\ba
\label{ei1}
\overline{({\rm Tr} (A \rho(t))^m} = \lim_{T \to \infty} \frac{1}{T}
  \int_0^T {\rm d} t \ ({\rm Tr} (A \rho(t))^m \ .
\ea
For $m = 1$ the long-time average gives the mean value of ${\rm Tr} (A
\rho(t))$, and powers of higher order are used to determine the
fluctuations of ${\rm Tr} (A \rho(t))$. We investigate that approach
by applying it to ${\rm Tr} (A \rho(t))$ defined in Eq.~(\ref{d8}) in
terms of the manifestly chaotic Hamiltonian~(\ref{d4}), and by
comparing the results with those of the random-matrix approach. We do
so first for the mean values and then for the time dependence of
thermalization.

In Ref.~\cite{Sre99}, the long-time average of ${\rm Tr} (A \rho(t))$
is shown to be equal to the thermal average of $A$ at the appropriate
temperature provided that the energy spread of the system is
sufficiently small compared to the mean energy. That constrains the
statistical operator $\Pi$. For the random Hamiltonian~(\ref{d4}),
calculation of the long-time average of ${\rm Tr} (A \rho(t))$ in
Eq.~(\ref{d8}) gives
\ba
\label{ei2}
\sum_{\mu \nu \rho \sigma \alpha} A_{\mu \nu} O_{\nu \alpha} O_{\rho \alpha}
\Pi_{\rho \sigma} O_{\sigma \alpha} O_{\mu \alpha} \ .
\ea
Averaging over the random matrix elements $O_{\mu \alpha}$ yields
three terms, two of which are of order $1 / N$. With ${\rm Tr} (\Pi) =
1$ the leading-order term is $(1 / N) {\rm Tr} (A)$. That agrees with
the first term on the right-hand side of Eq.~(\ref{s1}) and, thus,
with ${\rm Tr} (A \rho_{\rm eq}(\infty)$, in agreement with the result
of Ref.~\cite{Sre99}. (Actually, in the random-matrix approach there
is no need for any constraint on $\Pi$ since Eq.~(\ref{e2}) is an
immediate consequence of orthogonal invariance.) We note that in both
approaches, the average terms are actually present for all times $t >
0$. In our case that is seen directly from Eq.~(\ref{e2}). For
Ref.~\cite{Sre99} it follows from the fact that fluctuations around
the long-time average are calculated as fluctuations of (${\rm Tr} (A
\rho(t)) - {\rm Tr} (A \rho_{\rm eq})$).

We turn to thermalization. In Ref.~\cite{Sre99} the time dependence of
thermalization is determined in two steps. In the first step, higher
moments of (${\rm Tr} (A \rho(t)) - {\rm Tr} (A \rho_{\rm eq})$) are
estimated with the help of Eq.~(\ref{i1}). It is shown that in the
semiclassical regime and provided that the spread in energy of the
system is sufficiently small, the time evolution of (${\rm Tr} (A
\rho(t)) - {\rm Tr} (A \rho_{\rm eq})$) is with overwhelming
probability given by $C(t) {\rm Tr} (A \rho(0))$. Here $C(t)$ is
independent of the initial value ${\rm Tr} (A \rho(0))$ and is given
by
\ba
\label{ei3}
C(t) = \frac{\overline{{\rm Tr} (A \rho(t + t')) {\rm Tr} (A \rho(t'))}}
{\overline{{\rm Tr} (A \rho(t'))^2}}
\ea
where the long-time average is over $t'$. In the second step, the
numerator on the right-hand side of Eq.~(\ref{ei3}) is written in the
eigenstate representation of the Hamiltonian as $\sum_{\alpha \beta}
(A_{\alpha \beta})^2 (\Pi_{\beta \alpha})^2 \exp \{ i (E_\alpha -
E_\beta) t / \hbar \}$. Replacing $A_{\alpha \beta}$ by the last term
in Eq.~(\ref{i1}) and averaging over the ensemble results in
\ba
\label{ei5}
C_{\rm eth}(t) \propto \int_{- \infty}^{+ \infty} {\rm d} \omega \ | F(E,
\omega)|^2 \exp \{ i \omega t / \hbar \} \ ,
\ea
with $F(E, \omega)$ introduced in Eq.~(\ref{i1}). The index ``eth''
stands for the eigenstate thermalization hypothesis. As mentioned
earlier, in deriving Eq.~(\ref{ei5}) it is assumed that the
eigenvalues $E_\alpha$ are ordinary nonstochastic variables. If the
$E_\alpha$ were taken as random variables, one would have to average
$\sum_{\alpha \beta} (A_{\alpha \beta})^2 (\Pi_{\beta \alpha})^2 \exp
\{ i (E_\alpha - E_\beta) t / \hbar \}$ over the joint probability
distribution of $E_\alpha$ and $E_\beta$. Averaging would have to take
into account that the random variables $E_\alpha$ and $E_\beta$ occur
not only in the exponential but also as arguments of the function
$F(E, \omega)$ in Eq.~(\ref{i1}). The result would differ from
Eq.~(\ref{ei5}).

We compare the first step taken in Ref.~\cite{Sre99} with the
random-matrix model. There it is obvious almost from the outset (see
Eq.~(\ref{e2})) that the time dependence of thermalization is
independent of the initial value of $A$ and given by $\big\langle
|f(t)|^2 \big \rangle$. To compare with that result, we calculate
$C(t)$ from Eq.~(\ref{ei3}) and use Eq.~(\ref{d8}). We focus attention
on the numerator on the right-hand side of Eq.~(\ref{ei3}). In
averaging over the matrix elements $O_{\mu \alpha}$ we take into
account only contraction patterns that retain the dependence on
$t$. That is consistent with Ref.~\cite{Sre99}. Using $C(0) = 1$ we
find
\ba
\label{ei4}
C_{\rm rm}(t) = \big\langle |f(t)|^2 \big\rangle \ ,
\ea
where the index ``rm'' stands for the random-matrix
model. Eq.~(\ref{ei4}) is in agreement with Eq.~(\ref{s1}), confirming
the definition Eq.~(\ref{ei3}) of $C(t)$ given in Ref.~\cite{Sre99}
and the claim that the time evolution of thermalization is given by
$C(t) {\rm Tr} (A \rho(0))$.

The second step taken in Ref.~\cite{Sre99} leads to the explicit
expression for $C_{\rm eth}(t)$ in Eq.~(\ref{ei5}). That expression
differs from $C_{\rm rm}(t)$ in Eq.~(\ref{ei4}). The function $C_{\rm
  eth}(t)$ in Eq.~(\ref{ei5}) is the Fourier transform of the
non-negative function $|F(E, \omega)|^2$. Therefore, $C_{\rm eth}(t)$
oscillates in time about the value zero. If the band width in $\omega$
of $|F(E, \omega)|^2$ is finite, $C_{\rm eth}(t)$ is asymptotically
inversely proportional to $t$. In contrast, the function $C_{\rm
  rm}(t)$ in Eq.~(\ref{ei4}) is non-negative for all times $t$. To
leading order in $1 / N$, $\big \langle |f(t)|^2 \big\rangle$ is equal
to the square of the Fourier transform $g(t / \tau_\lambda)$ in
Eq.~(\ref{e10}) of the average level density and, for large $|t|$,
falls off as $1 / t^2$. Thus, the time dependence of thermalization is
fully determined analytically within the random-matrix model. Except
for the assumption on smoothness, the dependence on $E$ and $\omega$
of the function $F(E, \omega)$ in Eq.~(\ref{i1}) is not defined nor is
it obvious how information on that function might be
obtained. Actually, in Ref.~\cite{Sre99} the band width of the
function $F(E, \omega)$ with respect to $\omega$ is used as a
parameter and is estimated with the help of physical arguments.

The fundamental difference between the random-matrix model and the
eigenstate thermalization hypothesis lies in the identification of the
source of randomness. In the random-matrix model, it is the operator
$U(t)$ defined in Eq.~(\ref{d6}) and appearing in Eq.~(\ref{d8}) that
carries the random variables and that causes the function ${\rm Tr} (A
\rho(t))$ to be a stochastic process. The operator $A$, in
contradistinction, is written in the fixed basis of states $(\mu,
\nu)$ as in Eq.~(\ref{d7}) and, therefore, is not random. The central
result, Eq.~(\ref{e2}), is valid for any operator $A$. The approach of
Ref.~\cite{Sre99} takes the converse point of view. The elements
$\rho_{\alpha \beta}(t)$ of the density matrix of the system are
treated as ordinary (non-statistical) variables. The stochasticity of
the chaotic system is contained entirely in the eigenstate
representation $A_{\alpha \beta}$ of the operator $A$.

We see that Eq.~(\ref{i1}) is taylored to yield thermalization. From
the point of view of formal logic, Eq.~(\ref{i1}) formulates a
sufficient condition for thermalization. It is the great merit of
Ref.~\cite{Sre99} to have in that way given a first analytical
approach to and understanding of thermalization. A number of questions
remains. Why is randomness attached only to the elements $A_{\alpha
  \beta}$ of the operator $A$? Does Eq.~(\ref{i1}) impose a constraint
on the operator $A$? If so, which operators obey that constraint? Do
all operators that obey the constraint possess the same function $F(E,
\omega)$, or does that function depend on the operator $A$ under
consideration? How does $F(E, \omega)$ (beyond being smooth) actually
depend on the variables $E$ and $\omega$? The answers should link
Eq.~(\ref{i1}) with the Hamiltonian of the chaotic quantum
system. Without that link, Eq.~(\ref{i1}) remains a hypothesis.

Quantum chaos and stochasticity of the Hamiltonian are closely
linked~\cite{Boh84}. Therefore, it would seem more natural to analyze
thermalization in terms of a stochastic model for the Hamiltonian
rather than with the help of Eq.~(\ref{i1}). The random-matrix model
defined in Section~\ref{def} is too restricted for that purpose. The
approach of Ref.~\cite{Deu91} would seem a suitable
candidate. Unfortunately and in contrast to Ref.~\cite{Sre99}, it does
not offer the chance of an analytical treatment. That is the strength
of the eigenstate thermalization hypothesis.

\section{Discussion and Conclusions}
\label{dis}

We have shown that in the random-matrix model, every test function
${\rm Tr} (A \rho(t))$ thermalizes, i.e., attains the limit~(\ref{s1})
for $N \to \infty$. Thermalization is due to the orthogonal invariance
of the GOE: Orthogonal invariance causes the matrix elements $O_{\mu
  \alpha}$ in Eq.~(\ref{d4}) to be zero-centered Gaussian random
variables that obey Eq.~(\ref{d5}). That is the only fact used in
deriving Eq.~(\ref{e2}). In particular, orthogonal invariance
determines the form of the equilibrium term $(1 /N) {\rm Tr} (A)$. The
last term in Eq.~(\ref{s1}) gives the time dependence of
thermalization.  One might have expected that the time dependence of
thermalization is determined by destructive interference of the phase
factors $\exp \{ \pm i E_\alpha t / \hbar \}$ in Eq.~(\ref{i2}) for
the density matrix. That is not the case. The leading-order
contribution to thermalization is given by the function $\big\langle
f(t) \big\rangle$, the Fourier transform $g$ of the average GOE level
density. That function does not contain interference terms. The GOE
two-point level correlation function only yields a term of order $1 /
N$. Both functions are determined by orthogonal
invariance. Corresponding statements hold for the GUE where unitary
invariance takes the role of orthogonal invariance.

Thermalization as defined in Section~\ref{int} is a property of the
test function(s) ${\rm Tr} (A \rho(t))$. It characterizes the system
but is not an obvious property of the density matrix $\rho(t)$ in
Eq.~(\ref{i2}) itself. Thermalization is caused by the statistical
properties of the time-evolution operator $U(t)$ in
Eq.~(\ref{d8}). Thermalization is symmetric in time. That symmetry
holds for both, the time-reveral invariant GOE and the time-reversal
non-invariant GUE.

Comparison with the eigenstate thermalization hypothesis
Eq.~(\ref{i1}) shows that long-time averages agree with random-matrix
results. The predicted time dependence of thermalization is
qualitatively correct. The statistical assumptions used in
Ref.~\cite{Sre99} need justification. It would be desirable to link
form of and parameters in Eq.~(\ref{i1}) to the Hamiltonian of the
chaotic system.

Thermalization is very different from equilibration in open quantum
systems~\cite{Haa73}. The time dependence of equilibration is
described by a master equation that violates time-reversal
invariance. The statistical operator of the system changes with time
and tends toward the statistical equilibrium distribution. The process
is exponential in time. In contrast, thermalization occurs in an
isolated system. The statistical operator of the system does not
change with time, the elements of the density matrix of the system
keep oscillating in time forever. Only the test function ${\rm Tr} (A
\rho(t))$ thermalizes with a characteristic dependence on time $t$
given by $1 / t^2$.

In Ref.~\cite{Ued20}, the eigenstate thermalization hypothesis is
formulated without reference to Eq.~(\ref{i1}) and in more general
terms by stating that ``every eigenstate in an energy shell represents
thermal equilibrium''. We believe that our random-matrix models are
paradigmatic for the eigenstate thermalization hypothesis in that more
general form. The energy shell is the entire Hilbert space. All
eigenstates $| \alpha \rangle = \sum_\mu O_{\mu \alpha} | \mu \rangle$
are random superpositions of the basis states $| \mu \rangle$. The
statistical weight of every component vector $| \mu \rangle$ is the
same, independent of the eigenvalue $E_\alpha$ to which the state
belongs.

Thermalization tests the hypothesis that in an isolated system and in
the long-time limit, expectation values of operators tend to their
equilibrium values. That hypothesis is seen as the quantum analogue of
the ergodic hypothesis in classical statistical
mechanics~\cite{Dal16}. It is not surprising that in the random-matrix
models, thermalization follows from orthogonal or unitary
invariance. Such invariance implies that all states in $N$-dimensional
matrix space are equivalent. That is similar in spirit to classical
ergodicity.

{\bf Acknowledgement.} The author is grateful to T. Seligman for a
discussion and a suggestion.

\section*{Appendix~1: Low moments of the time-evolution operator $U$}

The matrix $U(t)$, defined in Eq.~(\ref{d6}), depends on the random
variables $O_{\mu \alpha}$ and $E_\alpha$. The former are Gaussian and
obey Eq.~(\ref{d5}). The latter obey Wigner-Dyson statistics. We
calculate the moments as averages over the matrix elements $O_{\mu
  \alpha}$. We do not, at this stage, average over the eigenvalues
$E_\alpha$. Averaging over the Gaussian-distributed matrix elements
$O_{\mu \alpha}$ is indicated by the index $O$ and is done by applying
Eq.~(\ref{d5}) to all pairs of matrix elements (``contracting'' pairs
of matrix elements). Our averaging procedure is justified in
Ref.~\cite{Pro02}. The ensemble average of $U(t)$ is
\ba
\label{a3}
\big\langle U_{\mu \nu}(t) \big\rangle_O = \delta_{\mu \nu} f(t)
\ea
where $f(t)$ is defined in Eq.~(\ref{e3}). The correlated part of the
second moment of $U(t)$ is
\ba
\label{a5}
\big\langle U_{\mu \nu}(t) U_{\mu' \nu'}(t) \big\rangle_{O, \rm corr} &=&
\frac{1}{N} (\delta_{\mu \mu'} \delta_{\nu \nu'} + \delta_{\mu \nu'}
\delta_{\nu \mu'}) f(2 t) \ , \nonumber \\
\big\langle U_{\mu \nu}(t) U^*_{\mu' \nu'}(t) \big\rangle_{\rm corr} &=&
\frac{1}{N} (\delta_{\mu \mu'} \delta_{\nu \nu'} + \delta_{\mu \nu'}
\delta_{\nu \mu'}) \ .
\ea
For the totally correlated part of the third moment of $U(t)$ we need
only the contribution that contains two factors $U(t)$ and one factor
$U^*(t)$. That term is given by
\ba
\label{a6}
&& \big\langle U_{\mu_1 \nu_1}(t) U_{\mu_2 \nu_2}(t) U^*_{\mu_3 \nu_3}(t)
\big\rangle_{O, \rm corr} = \frac{1}{N^2}  f(t) 
\nonumber \\
&& \qquad \times \big( \delta_{\nu_1 \mu_2} \delta_{\nu_2 \mu_3}
\delta_{\nu_3 \mu_1} + \ldots \big) \ .
\ea
The dots indicate the remaining seven possibilities to contract pairs
of matrix elements so as to yield the totally correlated part. For the
totally correlated part of the fourth moment of $U(t)$ we need only
the part that is bilinear in both $U(t)$ and in $U^*(t)$. That part is
given by
\ba
\label{a7}
&& \big\langle U_{\mu_1 \nu_1}(t) U_{\mu_2 \nu_2}(t) U^*_{\mu_3 \nu_3}(t)
U^*_{\mu_4 \nu_4}(t) \big\rangle_{O, \rm corr} = \frac{1}{N^3} \nonumber \\
&& \qquad \times \big( \delta_{\nu_1 \mu_2} \delta_{\nu_2 \mu_3}
\delta_{\nu_3 \mu_4} \delta_{\nu_4 \mu_1} + \ldots \big) \ .
\ea
The dots indicate the remaining possibilities to contract pairs of
matrix elements so as to yield the totally correlated part of the
fourth moment.

\section*{Appendix~2: Time Dependence of Thermalization}

For the uncorrelated part in Eq.~(\ref{e4}) we write the function
$f(t)$, defined in Eq.~(\ref{e3}), as
\ba
\label{e6}
f(t) = \frac{1}{N} \int {\rm d} E \ r(E) \exp \{ - i E t / \hbar \}
\ea
where
\ba
\label{e5}
r(E) = \sum_\alpha \delta(E - E_\alpha)
\ea
is the non-averaged GOE level density normalized to the total
number $N$ of states. The ensemble average of $r(E)$ is~\cite{Meh04}
\ba
\label{e7}
\langle r(E) \rangle = \frac{N}{\pi \lambda} \sqrt{1 - (E / (2
  \lambda)^2)} \ ,
\ea
where the range of $E$ is $- 2 \lambda \leq E \leq 2 \lambda$. That
gives
\ba
\label{e8}
\langle f(t) \rangle = \frac{1}{\pi \lambda} \int_{- 2 \lambda}^{2 \lambda}
        {\rm d} E \ \sqrt{1 - (E / (2 \lambda)^2} \exp \{ - i E t /
        \hbar \} \ .
\ea
Thus, $\langle f(t) \rangle$ is the Fourier transform of the average
GOE level density. We substitute $x = E / (2 \lambda)$, use
Eq.~(\ref{e9a}), and obtain
\ba
\label{b1}
\langle f(t) \rangle = g(t / \tau_\lambda) = \frac{4}{\pi} \int_0^1
{\rm d} x \ \sqrt{1 - x^2} \cos( x t / \tau_\lambda)
\ea
That is Eq.~(\ref{e10}). We cannot evaluate $g(t / \tau_\lambda)$
analytically for $t \neq 0$. An approximation to $g(t / \tau_\lambda)$
is obtained by replacing the square root function under the integral
by unity. Then $g(t / \tau_\lambda) = (4 / \pi) \sin (t /
\tau_\lambda) / (t / \tau_\lambda)$. That overestimates the correct
value $g(0) = 1$ by the factor $4 / \pi$. Numerical calculations
confirm the semiquantitative estimate. They show that a plot of $g(t /
\tau_\lambda)$ is similar in shape to that of the spherical Bessel
function $\sin (t / \tau_\lambda) / (t / \tau_\lambda)$, starting at
$g(0) = 1$, and oscillating about the value zero with decreasing
amplitude. The zeros are placed a distance $\approx \pi$ apart. To
estimate $g$ for large values of the argument, we divide the
integration interval in Eq.~(\ref{e10}) into $M$ equal sections of
length $1 / M$ each. We assume that $M$ is sufficiently large so that
in each such interval the square root function may be taken as
constant. It is then easy to show that $|g(t / \tau_\lambda)| < 2
\tau_\lambda / t$ for large $t$.

The correlated part of $f(t)$, written in full generality as a
function of two time arguments $(t_1, t_2)$, is given by
\ba
\label{e11}
&& \langle f(t_1) f^*(t_2) \rangle_{\rm corr} = \frac{1}{N^2}
\int_{- 2 \lambda}^{+ 2 \lambda} {\rm d} E_1 \int_{- 2 \lambda}^{+ 2 \lambda}
    {\rm d} E_2 \nonumber \\
&& \ \ \times \langle r(E_1) r(E_2) \rangle_{\rm corr} \exp \{ - i (E_1
t_1 - E_2 t_2) / \hbar \} \ .
\ea
For $N \gg 1$ the correlated part of the GOE level density is given
by~\cite{Meh04}
\ba
\label{e12}
\langle r(E_1) r(E_2) \rangle_{\rm corr} = - \langle r(E_1) \rangle \langle
r(E_2) \rangle Y_2(y) \ . 
\ea
Here $Y_2(y)$ with $y = (E_1 - E_2) / d$ is the two-level correlation
function of the GOE, and $d$ is the mean GOE level spacing. With $s(y)
= \sin(\pi y) / (\pi y)$, the function $Y_2$ is given by
\ba
\label{e12a}
Y_2(y) = s^2(y) + \bigg( \int_y^\infty {\rm d} t \ s(t) \bigg) \bigg(
\frac{\rm d}{{\rm d} y} s(y) \bigg) \ .
\ea
We define new integration variables $E = (1 / 2)(E_1 + E_2)$, $\ve =
E_1 - E_2$. Then $y = \ve / d$. The two-point function $Y_2$
effectively confines $y$ to values $y \ll N$ and, thus, $\ve$ to
values of order $d \ll 2 \lambda$. With $E_1 = E + (1/2) \ve$, $E_2 =
E - (1/2) \ve$, Eq.~(\ref{e7}) shows that the dependence on $\ve$ of
$\langle r(E_1) \rangle$ and of $\langle r(E_2) \rangle$ in
Eq.~(\ref{e11}) may be neglected. That gives
\ba
\label{e13}
&& \langle f(t_1) f^*(t_2) \rangle_{\rm corr} \\
&& = \frac{(- 1)}{(\pi \lambda)^2} \int_{- \infty}^{+ \infty} {\rm d} \ve
\ Y_2(\ve / d) \exp \{ - (i/2) \ve (t_1 + t_2) / \hbar \}\nonumber \\
&& \times \int_{- 2 \lambda}^{2 \lambda} {\rm d} E \ [1 - (E/(2 \lambda))^2]
\exp \{ - i (E (t_1 - t_2) / \hbar \} \ . \nonumber
\ea
We define $x = E / (2 \lambda)$, $\pi \lambda = N d$, use the
definition~(\ref{e9a}), and carry out the integral in the last
line. We obtain
\ba
\label{e15}
&& \langle f(t_1) f^*(t_2) \rangle_{\rm corr} \nonumber \\
&& = \frac{8}{\pi} \bigg(
\frac{\cos((t_1 - t_2) / \tau_\lambda)}{((t_1 - t_2) / \tau_\lambda)^2}
- \frac{\sin((t_1 - t_2) / \tau_\lambda)}{((t_1 - t_2) / \tau_\lambda)^3}
\bigg) \nonumber \\
&& \times \frac{1}{N} \int_{- \infty}^{+ \infty} {\rm d} y \ Y_2(y)
\cos (y (t_1 + t_2) / ( 2 \tau_d) \} \ .
\ea
The factor $1 / N$ is due to the substitution $\ve / (\pi \lambda) \to
y / N$. It accounts for the fact that the two-point correlation
function of the GOE is essentially confined to distances in energy
that are of order $d \ll 2 \lambda$. The time-dependent functions in
the second (third) line of Eq.~(\ref{e15}) display oscillations on the
time scale $\tau_\lambda$ ($\tau_d$, respectively). These are the
smallest (largest) time scales of the GOE. For $N \gg 1$ the
oscillations of the function in the third line are extremely slow
compared to those of the function in the second line. Putting $t_1 = t
= t_2$ we obtain Eq.~(\ref{e16}).


\end{document}